\documentclass[12pt]{article}

\usepackage[numbers,square,sort&compress]{natbib}
\usepackage{url}
\usepackage{breakurl}

\usepackage[breaklinks]{hyperref}
\usepackage{xcolor}
\usepackage[margin=2.5cm]{geometry}
\hypersetup{
    colorlinks,
    linkcolor={red!50!black},
    citecolor={blue!50!black},
    urlcolor={blue!80!black}
}
\usepackage{graphicx}
\usepackage{amsmath}
\usepackage{ifthen}

\newif\ifrevision
\revisionfalse

\newif\ifrebuttal
\rebuttalfalse

\ifrevision

\else

\fi

\newcommand{\beginsupplement}{%
        \setcounter{table}{0}
        \renewcommand{\thetable}{S\arabic{table}}%
        \setcounter{figure}{0}
        \renewcommand{\thefigure}{S\arabic{figure}}%
        \setcounter{equation}{0}
        \renewcommand{\theequation}{S\arabic{equation}}%
     }

\newcommand{\fig}[3][1]{
    \begin{figure}
        \centering
        \includegraphics[width=#1\textwidth]{fig/#2}
        \caption{#3 \label{fig/#2}}
    \end{figure}
    }

\newcommand{\hfig}[3][1]{
    \begin{figure}[h]
        \centering
        \includegraphics[width=#1\textwidth]{fig/#2}
        \caption{#3 \label{fig/#2}}
    \end{figure}
    }
    
\newcommand{\figref}[2]{Figure\,\ref{fig/#1}{#2}}

\newcommand{\tab}[5]{ 
    \begin{table}     
        \centering
        \caption{
            #2 \label{tab/#1}
        }    
        \begin{tabular}{#3} 
        \hline 
            #4 \\
        \hline
         #5 
        \end{tabular}
    \end{table}
}

\newcommand{\tabref}[1]{Table\,\ref{tab/#1}}

\newcommand{\eq}[2]{
    \begin{equation}
        #2 \label{#1}
    \end{equation}
}

\newcommand{\eqarr}[1]{
    \begin{eqnarray}
        #1
    \end{eqnarray}
}

\renewcommand{\(}{\left(}
\renewcommand{\)}{\right)}
\renewcommand{\[}{\left[}
\renewcommand{\]}{\right]}

\newcommand{\til}{\tilde}

\newcommand{\al}{\ensuremath{\alpha}}

\newcommand{\be}{\ensuremath{\beta}}

\newcommand{\ga}{\ensuremath{\gamma}}

\newcommand{\ka}{\ensuremath{\kappa}}

\newcommand{\la}{\ensuremath{\lambda}}

\newcommand{\rh}{\ensuremath{\rho}}

\newcommand{\si}{\ensuremath{\sigma}}

\newcommand{\ch}{\ensuremath{\chi}}

\title{
    An autonomous compartmental model for accelerating epidemics
}

\author{
Nazmi Burak Budanur\thanks{nbudanur@pks.mpg.de} $^{1,2}$ and Björn Hof $^2$ \\ 
$^1$\small{Max Planck Institute for the Physics of Complex Systems (MPIPKS)}\\
    \small{Nöthnitzer Straße 38, 01187 Dresden, Germany} \\
$^2$\small{Institute of Science and Technology Austria (IST Austria)} \\ 
    \small{Am Campus 1, 3400 Klosterneuburg, Austria}
}

\date{\today}

\begin{document}

\maketitle 

\abstract{
    In Fall 2020, several European countries reported rapid increases in  
    COVID-19 cases along with growing estimates of the effective reproduction
    rates. Such an acceleration in epidemic spread is usually attributed to
    time-dependent effects, e.g. human travel, seasonal behavioral changes,
    mutations of the pathogen etc. In this case however the acceleration
    occurred when counter measures such as testing and contact tracing exceeded
    their capacity limit. Considering Austria as an example, here we show that
    this dynamics can be captured by a time-independent, i.e. \emph{autonomous},
    compartmental model that incorporates these capacity limits. In this model,
    the epidemic acceleration coincides with the exhaustion of mitigation
    efforts, resulting in an increasing fraction of undetected cases that drive
    the effective reproduction rate progressively higher. We demonstrate that
    standard models which does not include this effect necessarily result in a
    systematic underestimation of the effective reproduction rate.}

\section{Introduction}
\label{introduction}

Severe acute respiratory syndrome coronavirus 2 (SARS‑CoV‑2)
\cite{bar-on2020sarscov2} and its variants \cite{who2021variants} continue to
challenge our lives \cite{hale2021global} even after the rollout of several
effective vaccines \cite{mathieu2021global}. As of Fall 2021,
non-pharmaceutical interventions such as mask mandates and travel restrictions
at varying degrees still remain in place around the globe \cite{hale2021global}.
These interventions are guided by almost-real-time assessment of the ongoing
epidemiological situation which rely on surveillance data and mathematical
models and, are thus, prone to their uncertainties and shortcomings
\cite{manrubia2020uncertain}. It is, therefore, crucial for decision making that
the epidemiological models are sufficiently \emph{simple} to be used in a 
fast-changing environment while containing the necessary amount of \emph{complexity} 
to capture all essential features of the real epidemic. 

Simple epidemic models divide a population into ``compartments'' according to
individuals' epidemiological status and specify the rules by which the disease
progresses within an individual and spreads over the population
\cite{keeling2008modeling}. In the most basic form, these rules are given as
transition rates between the compartments which can be translated into a set of
ordinary differential equations (ODEs). One such model is the SEIR model where
the compartments correspond to those who are  
susceptible (S) to infection, exposed (E) to the pathogen (but not yet a
spreader), infectious (I), and removed (R) from epidemic dynamics (dead or
immune). The SEIR model can capture the initial exponential increase of
infections when the majority of a population is susceptible and the subsequent
slow down of spreading due to the continuously increasing removed population. 

In Fall 2020, the second wave of the COVID-19 in Austria exhibited a remarkably
different behavior than the one that we described above. What appeared to be
slow-but-steady initial increase was followed by an \emph{acceleration} of the
epidemic indicated by a faster-than-exponential growth in case numbers
\cite{taschwer2020woran}. For simple mathematical models, such an observation
indicates a change in rules, i.e. increased transmissibility, which can be due
to seasonality \cite{merow2020seasonality,liu2021role} or occurrence of more
infectious variants \cite{korber2020tracking}. In contrast to such expectations,
here we show that the accelerating increase of COVID-19 cases reported in
Austria during Fall 2020 can be captured in an autonomous compartmental model
described by a time-independent (deterministic) set of ODEs. As we explain in
the following, the key modeling ingredient for accelerating epidemics is
explicit inclusion of mitigation efforts and their capacity limitations within
the model dynamics. 

Since the early stages of the pandemic, case and contact isolation has been one
of the primary public health responses \cite{hellewell2020feasibility}. Using
stochastic agent-based epidemic models on networks,
\citet{scarselli2021discontinuous} showed that while testing and contact
isolation slow down an epidemic, limiting the number of available tests
fundamentally alter its nature by changing the epidemic transition for large
populations (in the thermodynamic limit) from gradual (second order) to sudden
(first order). The driving mechanism of this qualitative change was the
acceleration of transmissions when the models' testing capacity were exhausted.
In the present work, we introduce the \emph{SEIRTC model} which, in addition to
the $S$, $E$, $I$, and $R$, includes separate compartments for the tested
individuals (T) and confirmed cases (C) similar to
\cite{scarselli2021discontinuous}. Differently from the stochastic network
models used in \cite{scarselli2021discontinuous}, the dynamics of the SEIRTC
model is determined by a set of ODEs, rendering it suitable for working with
real data. 
    In other words, our aim in the present paper is to present a simple way 
    of incorporating capacity-limited interventions to the standard 
    epidemic models so that it can be utilized to explain the real-world 
    observations without resorting to complex network-based models. In this 
    sense, our approach is similar to that of Arino et al. 
    \cite{arino2006simple}, who presented an extension of the SEIR model that 
    includes treatment as a counter-acting measure to explain influenza data. 
We show that the SEIRTC model can be fitted to the COVID-19
surveillance data published by the Austrian Agency for Health and Food Safety
(AGES) \cite{ages} and capture the epidemic acceleration observed in Fall 2020
without the need for a temporal modification of the infectiousness. Our results
suggests that during this period, the effective reproduction rate, i.e. the
average number of secondary cases originating from a primary one after the 
initial uncontrolled spread period, was systematically underestimated. 

\section{Methods}
\label{methods}

We begin with a brief recapitulation of the standard SEIR model 
    which forms the basis of our SEIRTC model to follow. For a detailed 
    treatment, we refer the reader to 
    \cite{keeling2008modeling,hethcote2000mathematics}. The SEIR model is
represented by the state transition diagram in \figref{models}{A} and the
corresponding ODEs can be written by expressing the rates of changes in
compartments' populations according to the transition rates implied by the
annotated arrows as
\eq{SEIR}{
    \dot{S} = - \be I S / N ,\quad  
    \dot{E} = \be I S / N - \ga_E E ,\quad  
    \dot{I} = \ga_E E - \ga_I I ,\quad  
    \dot{R} = \ga_I I ,
}
where, $\dot{}$ denotes derivative with respect to time, $N = S + E + I + R$ is
the population, $\be$ is the transmission parameter, and $\ga_E$ and $\ga_I$ are 
the inverse latent and infectious times, respectively. 
The transmission parameter \be\ can be interpreted as the number of interactions 
per person per unit time multiplied by the transmission probability at each 
interaction \cite{keeling2008modeling}.
The underlying assumption of SEIR model is that the population 
is well mixed, thus, the effects solely due to the network heterogeneities are 
neglected. 

\fig{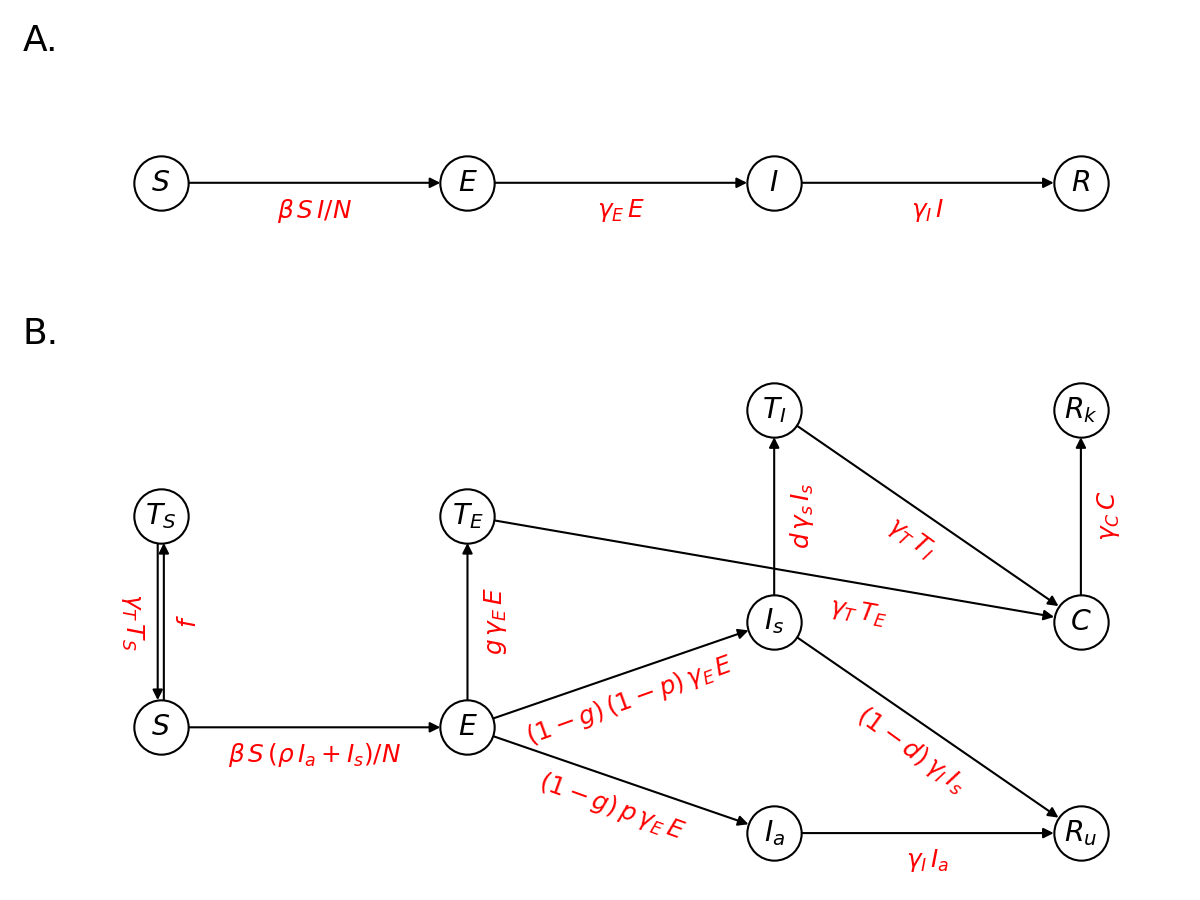}{
    State transition diagrams of SEIR (A) and SEIRTC (B) models where the
    encircled letters denote the compartments into which the population is
    divided and the arrows along with their labels underneath indicate the 
    transition rates between the compartments. The compartments are 
    $S$: susceptible, $E$: exposed, $I$: infectious, 
    $R$: removed (recovered or dead), 
    $I_{a/s}$: a/symptomatic infectious,  
    $R_{u/k}$: un/known removed,
    $T_{S/E/I}$: tested susceptible/exposed/infectious, 
    $C$: case.
}

    We obtain the SEIRTC model shown in \figref{models}{B} by the following series of 
    modifications to the SEIR. 
    First, we split the infectious individuals $I$ into two sub-compratments, 
    namely $I_s$ ``symptomatic'' and $I_a$ ``asymptomatic'',
the latter of whom are those who show no symptoms throughout their
infectious period and spread the disease at a relative risk \rh\ as implied by
the $S \rightarrow E$ term. At this stage, our model is equivalent to the SLIAR 
model of Arino et al. \cite{arino2006simple}.
Next, we incorporate testing into our model by introducing the 
compartments $T_S$, $T_E$, and $T_I$, where the subscripts refer to the 
epidemiological state of the individuals who are tested. We assume that 
testing with symptoms also invokes isolation, i.e. the individuals 
in the $T_I$ state can no longer spread the disease. 
The susceptibles who are tested ($T_S$) return to $S$, whereas the 
exposed and infectious individuals become ``cases'' ($C$) after being tested. 
Finally, the removed individuals are split into ``known'' ($R_k$) and 
``unknown'' ($R_u$) parts for convenience in presenting our results to follow.

Similar to those in the SEIR model, the parameters $\ga_x$ of the SEIRTC model
refer to the inverse mean lifetime at the compartment $x$, with the exception
$\ga_s$ which is the inverse mean symptom onset time, at which point a
symptomatic infectious individual becomes detectable. Hence, this delay term
accounts for presymptomatic transmissions which are believed to play a significant
role in spreading {COVID-19} \cite{arons2020presymptomatic}.  
$\ga_T$ is the test turn-over time, i.e. the time from the administration of a
test to its outcome. Finally, 
$\ga_C$ is not an independent parameter but is determined as
\eq{ga_C}{
    \ga_C = 
        \[\ga_I^{-1} 
        - \frac{T_E \ga_T^{-1} + T_I (\ga_s^{-1} + \ga_T^{-1})}{
                T_E + T_I}\]^{-1} , 
}
where we assume that at time $t$, the ratio of cases that are identified before
and after developing symptoms are proportional to the number of individuals in
$T_E$ and $T_I$, respectively. In doing so, we neglect a delay by
$\gamma_T^{-1}$ and avoid working with delay-differential equations 
which are complicated to work with numerically.

The transition rates in \figref{models}{B} have several probabilistic factors.
These are $p$: probability of an asymptomatic infection, $\rh$: relative risk of
transmission from an asymptomatic individual, $d$: probability of detecting 
a symptomatic infectious individual via testing, $g$: probability of 
detecting an exposed individual via contact tracing before becoming 
infectious. All but the last one of these
probabilities are independent parameters to be determined via literature
estimates or model fitting. Because the detection of an infection before
developing any symptoms can only be possible via contact tracing, the
probability $g$ is a complex function of the number of (un)identified infections, 
the underlying social network structure, and the contact tracing policy and 
capacity. 
    Since these details are not within the scope of the present model, here, we 
    resort to an ansatz that is based on two simplifying assumptions: 
    (i) The probability of detecting a case is proportional to the ratio 
    $C / (C + I_s + I_a)$ of known cases to those that are undetected at 
    a given time,
    (ii) Total contact tracing capacity is limited such that no more than 
    $T_m$ tests on susceptible and exposed individuals can be carried out at an 
    instance. Let $H(x)$ be the Heaviside step function that takes values 
    $H(x) = 0 \mbox{ for } x < 0$ and $H(x) = 1 \mbox{ for } x > 0$, 
\eq{g_ansatz}{
    g = \ka \frac{C}{C + I_s + I_a} H(T_m - T_S - T_E)
}
fulfils the assumptions that we stated above. We assume the rest of the factors 
such as the ratio of false negative tests and likelihood of contact tracing 
are averaged into the fit parameter $\ka$. In our 
implementation, we approximate the step function as 
$H(x) \approx 1/2 + (1/2) \tanh (x)$. 
Because contact tracing is only possible through known cases, we expect the 
the probability $g$ to increase with the ratio of identified cases 
to total number of infectious individuals and the ansatz \eqref{g_ansatz} 
should be understood as the simplest expression that agrees with this 
intuition. While opting for model simplicity, we neglect 
beyond-linear-order terms, such as those proportional $[C / (C+I_s+I_a)]^3$ and 
$[C / (C+I_s+I_a)]^5$, and delays since taking the 
probability of an exposed case to be detected at time $t$ to be a function of 
the number of cases and infectious individuals at time $t$ ignores the latent 
time $\ga_E$ from exposure to become detectable.

Similar to $g$, the number of susceptible individuals to be tested per unit 
time $f$ is also an unknown function of the social network and contact tracing 
procedures. Because in this case the individuals are not exposed to the pathogen, 
    we assume that this rate is independent of the number of infectious 
    individuals at present and  
take 
\eq{f_ansatz}{
    f = (\al S C / N) H(T_m - T_S - T_E) \,, 
    }
as our second ansatz, where the term $\al S C / N$ is analogous to the 
$S \rightarrow E$ term of the SEIR model (\figref{models}{A}).
    Here, we assume that the daily probability of a susceptible person 
    to have a past contact with a known case wherein no infection have occured 
    is proportional to the fit parameter $\al$ and is independent from the 
    probability of being infected at the same time, i.e. 
    the $S \rightarrow E$ transition. 
    Once again, the step function $H(T_m - T_S - T_E)$ approximated as 
    $H(x) \approx 1/2 + (1/2) \tanh (x)$ models the capacity limit of contact 
    tracing by setting $S \rightarrow T_S$ flux to $0$, once the capacity 
    limit $T_m$ is reached.
    Finally, we ignore false positives, and thus let 
    all individuals from the compartment $T_S$ back to $S$ after the turnover 
    time $1/\ga_T$.

As we illustrate through our results of the next section, 
the presence of $H(T_m - T_S - T_E)$ terms in \eqref{g_ansatz} and \eqref{f_ansatz} 
limits the total number of individuals in $T_S$ and $T_E$ compartments to 
$T_m$ by substantially reducing $S \rightarrow T_S$ and $E \rightarrow T_E$ fluxes, 
as this limit is approached. This bound is not imposed upon the testing of the 
symptomatic individuals, i.e. the $I_s \rightarrow T_s$ term, since we assume it 
to be due to contact tracing. Finally, we assume that if an asymptomatic individual 
is detected via contact-tracing, this takes place before the individual becomes 
infectious, which is implied by the fact that those in $I_a$ are not tested.  

With the ansätze \eqref{g_ansatz} and \eqref{f_ansatz}, the SEIRTC ODEs similar
to \eqref{SEIR} corresponding to the state transition diagram \figref{models}{B}
can be obtained by expressing the rates of change in compartment populations as
the shown transition rates. 
Explicit form of these equations (\ref{dotS}--\ref{dotR_k})
    can be found in the supplementary material.  
    In our numerical results to follow, we simulate
these using \texttt{odeint} function of \texttt{scipy} \cite{virtanen2020scipy}.

For model fitting and uncertainty quantification, we follow 
\cite{chowell2017fitting} and utilize weighted nonlinear least squares fit 
\cite{banks2014modeling} for adjusting model parameters followed by a 
bootstrap method \cite{efron1994introduction} for finding alternative sets 
of fit parameters. 
In a real-life scenario, testing constitutes the primary source of information
as most countries publish the daily numbers of tests they conduct and those with
a positive outcome (incidence). We make use of both of these
measurements and minimize the cost function 
\eq{cost_function}{J  =  
    \sum_{n = 0}^{N} \frac{1}{\til{T}} 
        \(\til{T} [n] - T'[n] \)^2  
    + \sum_{n = 1}^{N}  \frac{1}{\til{C}}
        \(\til{C} [n] -  C'[n]\)^2 \,, 
}
where $[n]$ denotes the discrete time in days, 
$\til{}$ indicates measurements coming from the surveillance data, and
$T'[n] = \sum_{i = S, E, I} T_i [n] (\gamma_T \times \mbox{day})$ and
$C'[n] = C [n] + R_k[n] - C[n - 1] - R_k [n - 1]$ are the number of tests 
carried out and
new cases 
recorded in the model on day $n$, 
    respectively. In the following, we take $1 / \gamma_T = 1\,\mbox{day}$, which 
    renders the factor $\gamma_T \times \mbox{days} = 1$, i.e. each individual  
    remains in the test compartment for 1 day. 
    Note that because our definition of daily new cases $C'[n]$ in the
    SEIRTC model depends on the total number of $C + R_k$ of the previous day, the
    corresponding term in the cost function start from the day $1$ rather than $0$. 
    The choice of weights $\til{T}^{-1} [n]$ and $\til{C}^{-1} [n]$ ensures that the 
    optimization algorithm does not ignore the earlier stages in favor of the later 
    days on which the case and test numbers are much higher.
In order to reduce the number of fit 
parameters, we make the following simplifying assumptions.
Whenever available, we take literature values for parameters or restrict them 
to the established estimated interval.  
While the number of exposed individuals on day 0 is varied, 
the initial populations of $T_S, T_E, I_a, I_s, T_I, C$ are 
adjusted through a fixed-point iteration that transfers individuals from $S$ 
to these compartments while minimizing the error between the simulated dynamics 
of the first 10 days from exponential fits. 
$R_k(0)$ and $R_u(0)$ are both set to the number of cases registered until the 
first day. Although this is an arbitrary assumption for $R_u$, it has no 
effect on the dynamics as long as it is much less than the total population, 
which is the case for our results to follow. 
With these assumptions, the set of fit parameters becomes  
$\theta = \{E(0), \al, \ka, T_m, \be, \ga_s, d \}$.
In our applications, we utilized \texttt{least\_squares} method of 
\texttt{scipy} \cite{virtanen2020scipy} to minimize 
\eqref{cost_function} and find the best-fit parameters $\theta^*$.
For bootstrapping, we generate synthetic data 
$\hat{T}[n] = \mathrm{Pois}(T'[n]), \hat{C}[n] = \mathrm{Pois}(C'[n])$, 
where $\mathrm{Pois}(\la)$ indicates a random variable drawn from a 
Poisson distribution \cite{papoulis1967probability} with the mean and 
variance $\la$, and refit our model by taking the 
$\hat{T}[n], \hat{C}[n]$ as our new set of observations in 
\eqref{cost_function}. This procedure is illustrated in 
\figref{bootstrap}{} of the supplement.

In the following for comparison, we also present fits by SEIR models for 
which we discard the test measurements from \eqref{cost_function} and take 
the cost function 
\eq{cost_function_SEIR}{
    J = \sum_{n = 1}^{N} w_C[n] \(\til{C} [n] -  C'[n]\)^2 \,, 
}
where $C'[n] = I[n] + R[n] - I[n-1] - R[n-1]$ in the SEIR model. 
Similar to the SEIRTC model, 
we take $E(0)$ and $\be$ as fit parameters and initiate simulations such 
that the dynamics of the first $10$ days can be approximated by exponentials. 
In order to illustrate how the SEIR model fits the different stages 
of the time-interval considered for different weights, we consider two 
different choice of weights, namely $w_C[n] = 1$ and $w_C[n] = 1 / \til{C}$. 

In order to estimate the effective reproduction number ($R_t$) from
surveillance and model data we utilize the \texttt{python} implementation
\cite{huisman2020estimation} of the \citet{cori2013new}'s \texttt{EpiEstim}
algorithm that is based on the Bayesian inference of $R_t$ from a
Gamma-distributed prior assuming Poisson-distributed transmissions. This method
was also used by AGES \cite{richter2020epidemiologische} who performs the 
real-time epidemilogical monitoring of the ongoing COVID-19 situation in 
Austria.

\section{Results}
\label{results}

\fig{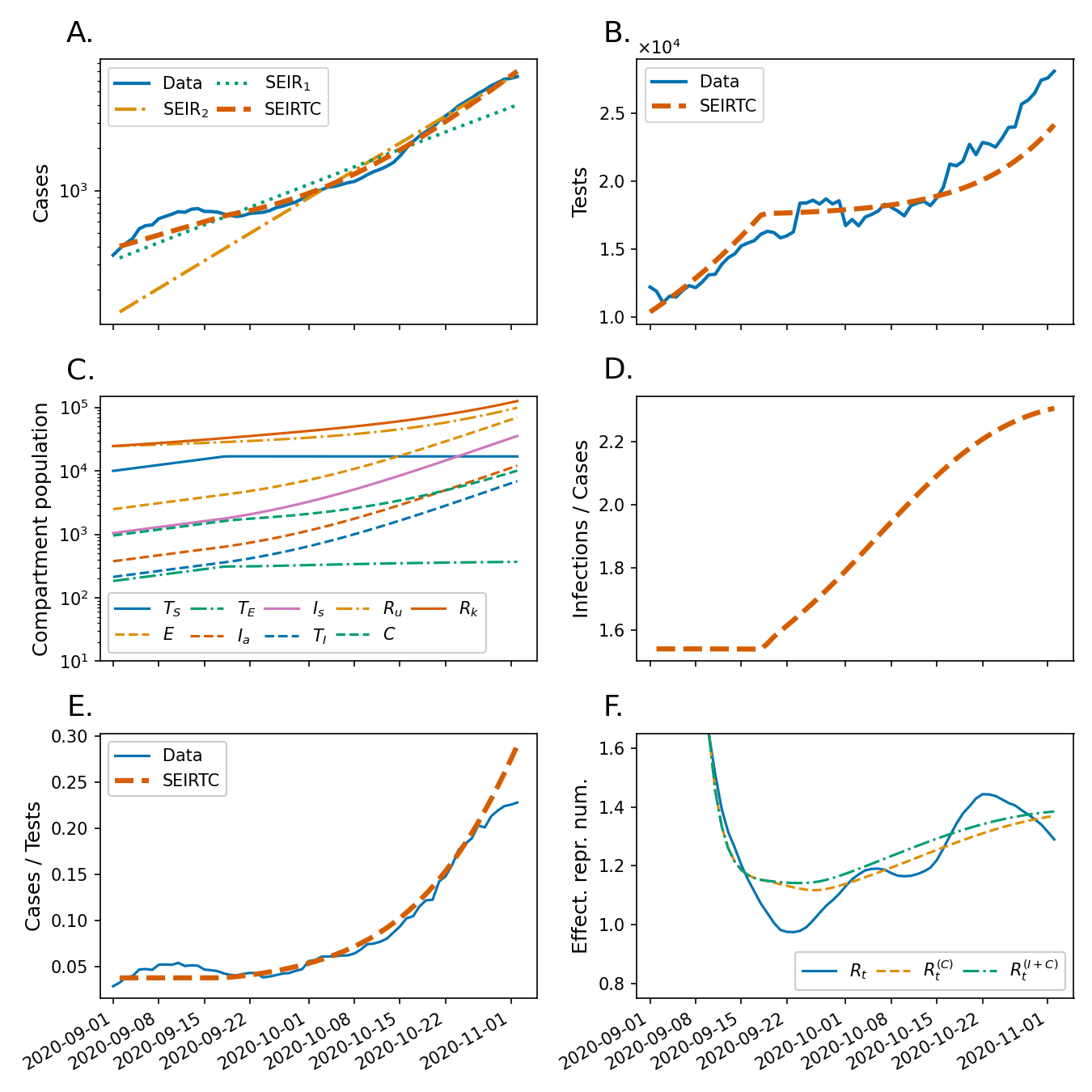}{
    Case ({\bf A}) and Test ({\bf B}) data (7-day moving average) reported in 
    Austria from September 1, 2020 to November 3, 2020 along with the fitting 
    curves obtained from the SEIRTC model. 
    Two SEIR model fits using different cost functions weights (see the main 
    text) are also shown in {\bf A} for comparison. 
    {\bf C.} Populations of the individual compartments (except $S$) 
    of the SEIRTC model.
    {\bf D.} The ratio of daily new infections to the cases  
    in the SEIRTC model.
    {\bf D.} The reported proportion of tests with a positive outcome and 
    that in the SEIRTC model.     
    {\bf F.} 
    Estimates $R_t$, $R_t^{(C)}$, and $R_t^{(I+C)}$ of the effective 
    reproduction number based on 
    the reported case data, 
    case numbers of the SEIRTC model, 
    and combined case infection numbers of, respectively. 
}

We consider the second wave of COVID-19 in Austria from September 1 to November 
3, 2020, on which day the country went into its second lockdown in 
order to protect its healthcare system from an otherwise-inevitable overload.  
\figref{fits}{A,\,B} shows the 7-day moving averages of the numbers of 
confirmed cases and performed tests during this period, respectively 
(retrived from \citep{ages}). Fits to these data by the SEIRTC model 
with parameters in \tabref{parameters} are also 
shown in \figref{fits}{A,\,B}, which we obtained by minimizing 
\eqref{cost_function}. 
For comparison in \figref{fits}{A}, we also show fits by
SEIR models with initial conditions and model parameters as listed in 
    \tabref{parameters_SEIR}. The different choice of weights in 
    \eqref{cost_function_SEIR} results in $SEIR$ models, with different set 
    of model parameters and initial conditions, which we refer to as  
    SEIR$_1$ and SEIR$_2$. As shown in \figref{fits}{A}, 
when unit weights are chosen, the SEIR$_2$ model  
underestimates the initial case numbers whereas when the weights are 
inversely proportional to the daily number of confirmed cases, the later case 
numbers are underestimated by the SEIR$_1$ model. 
In contrast, the fit by the SEIRTC model captures both episodes.
    This is further illustrated by the scatter plots of model predictions
    against the observations in \figref{fitscatter}{} where the largest 
    deviation between the two are also marked for each model. 
    Quantitatively, the largest percentage relative error between the case 
    numbers and their model predictions, i.e. 
    $100 \times \max_n (\til{C}[n] - C'[n]) / C'[n]$, are 
    $39.3\%$ and $68.8\%$ for SEIR$_1$ and SEIR$_2$ models, respectively, 
    whereas it is $25.0\%$ for the SEIRTC model. In addition, 
    we also carried out a reduced-$\ch^2$ ``goodness of fit'' 
    test taking our model predictions as means of Poisson distributions, 
    see the supplemental material for details. $\ch^2_\nu = \ch^2 / \nu$ where 
    $\nu$ is the number of degrees of freedom is 
    $\ch^2_\nu = 223.1$ and $\ch^2_\nu = 207.1$ for 
    SEIR$_1$ and SEIR$_2$ models, respectively, 
    whereas $\ch^2_\nu = 86.7$ for the SEIRTC model. The lower 
    $\ch^2_\nu$ of the SEIRTC model further demonstrates that it is a better 
    fit to the data considered \cite{barlow1993statistics}. 
    In addition to the goodness of fit test, we also performed a Poisson 
    bootstrap analysis to reveal parameter correlations of the SEIRTC model. 
    Although some of the model parameters show correlations 
    (see \figref{parameter_scatter}) our main results are robust to these 
    parameter variations.

\tab{parameters}{
    The model parameter values which are used in the fit shown in \figref{fits}. 
}{
    c c | c  c 
}
{
    Symbol      & Value [Ref.] / [Fit]                         
        & Symbol      & Value [Ref.] / [Fit] 
}{
    N           & $8\,894\,380$ \citep{auinger2021abgestimmte} 
        & \be         & 0.5993 / person / day [Fit] \\ 
    \al         & 10.84 / person / day [Fit]                   
        & \ka         & 0.7546 [Fit]  \\
    p           & 0.2 \citep{buitrago-garcia2020asymptomatic}   
        & d           & 0.5446 [Fit]  \\
    $1 / \ga_E$ & 4 days \citep{weitz2020modeling}             
        & $1 / \ga_T$ & 1 day \citep{aerztekammer} \\
    $1 / \ga_I$ & 5 days \citep{weitz2020modeling}             
        & $1 / \ga_s$ & 2.594 days [Fit] \\
    $T_m$ & 17\,252 [Fit]                                      
        & $\rh$       & 0.35 \citep{buitrago-garcia2020asymptomatic}
}

\tab{parameters_SEIR}{
        Cost function weights and parameters for SEIR models with fit curves 
        shown in \figref{fits}.
}{
    c c c c c c c c c  
}
{
    Model 
    & $w_C [n]$
    & $1 / \ga_E$
    & $\ga_I$
    & N
    & E(0)
    & I(0)
    & R(0)
    & $\be$
    }{
    SEIR$_1$
    & $1 / \til{C} [n]$
    & $4$ days \citep{weitz2020modeling}
    & $5$ days \citep{weitz2020modeling}
    & $8\,894\,380$ \citep{auinger2021abgestimmte}
    & 1306
    & 1353
    & 5390
    & 0.282 \\
      SEIR$_2$
    & $1$
    & $4$ days \citep{weitz2020modeling} 
    & $5$ days \citep{weitz2020modeling} 
    & $8\,894\,380$ \citep{auinger2021abgestimmte} 
    & 535
    & 506
    & 1518
    & 0.332
}

As shown in \figref{fits}{B}, the SEIRTC model also captures the 
reported test numbers with a visible change in its trend during the days 
following September 15. This coincides with the exhaustion of the contact 
tracing capacity as reflected by the plateaus of $T_S$ and $T_E$ in  
\figref{fits}{C}.  Consequently, the ratio $I'[n] / C'[n]$ 
(\figref{fits}{D}), where 
$I'[n] = \sum_i I_i[n] + \sum_i R_i[n] + C[n] 
- \sum_i I_i[n - 1] + \sum_i R_i[n - 1] + C[n - 1]$, of the daily new 
infections to those that are detected starts growing, yielding more and more 
infections that remain undetected. An observable consequence of this is the 
increase of the proportion of the tests with a positive outcome as shown in 
\figref{fits}{E} where we plotted this quantity using the data reported and 
those obtained from the SEIRTC model.

In \figref{fits}{F}, we show the estimates of the effective reproduction 
number $R_t$ calculated using the case data shown in \figref{fits}{A}
and those computed from the SEIRTC model's daily case 
(orange, dashed) and infections (green, dotted-dashed). 
In these calculations, we chose a smoothing window of 7 days and used a serial 
interval obtained by discretizing a Gamma distribution with a mean $4.46$ and 
a standard deviation $2.63$ days as estimated by AGES 
\cite{richter2020epidemiologische}. 

\fig{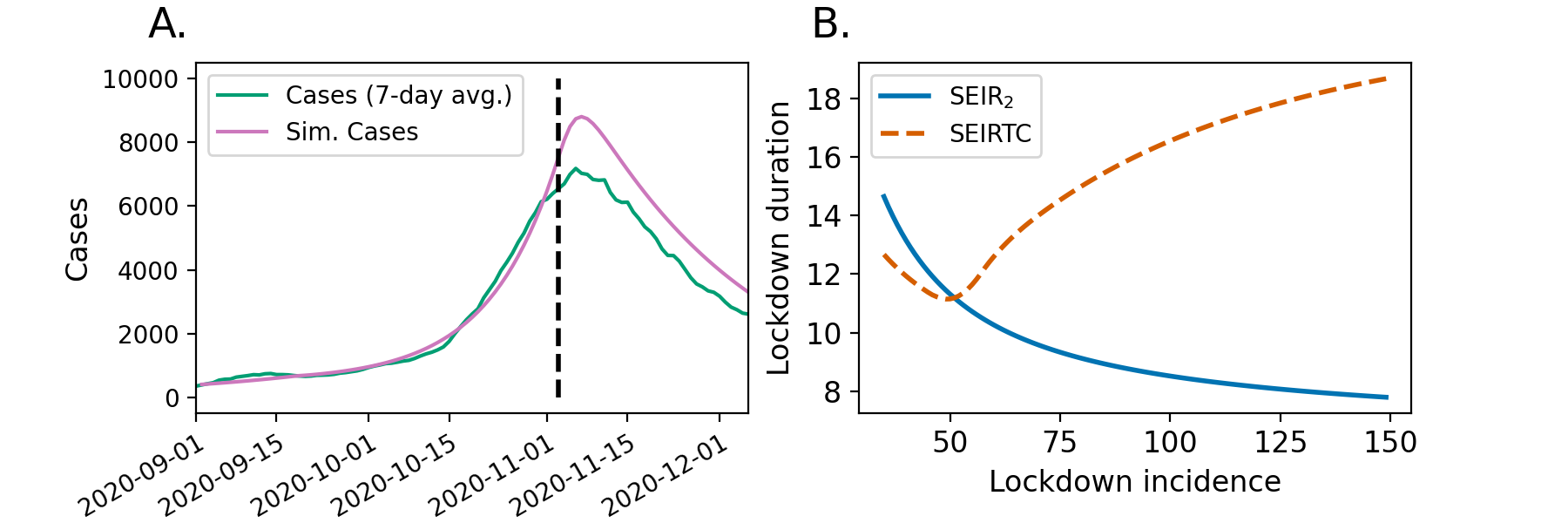}{
        \textbf{A.} Case data reported in Austria from 
        September 1, 2020 to December 6, 2020 along with its prediction by 
        SEIRTC model wherein the lockdown is modeled by a reduction of $\be$ 
        to $45\%$ of its value on November 3 which is indicated by the dashed
        line segment. \textbf{B.} 
        Number of days in ``lockdown'' necessary for reducing the
        7-day incidence (per 100,000) by 10 as a function of the incidence at which
        the lockdown begins. Lockdown is modeled by a drop of the transmission
        parameter $\be$ to half of its value.
    }

About a week after the beginning of lockdown in Austria on November 3 2020, 
the case numbers began to decrease as shown in \figref{lockdowns}{A}. We 
observed that this trend can be qualitatively captured by the SEIRTC model 
if the transmission parameter $\be$ is reduced to $45\%$ of its value on the 
same day as shown in \figref{lockdowns}{A}. Taking this as a crude model of 
a lockdown, we investigated hypothetical scenarios during which 
the transmission parameter $\beta$ is reduced to $0.45\be$ beginning 
at different times. Specifically, we started SEIR$_2$ and
SEIRTC models with their initial conditions and model parameters same as those
that we used in \figref{fits}, initiated lockdowns when the 7-day incidence
(per 100,000 people) reached a certain value, and measured the lockdown duration 
necessary for reducing the incidence by 10. The results of these simulations 
are shown in \figref{lockdowns}. 

\section{Discussion}
\label{discussion}

\figref{fits}{A,\,B} demonstrates that the SEIRTC model with parameters listed
in \tabref{parameters} and the initial population shown in \figref{fits}{C} is
able to capture the accelerating spread of COVID-19 observed in Austria in Fall
2020.
We note that some of these fit parameters are correlated, hence, it is
possible to find different set of parameters that result in comparable fitting
errors, see \figref{bootstrap}{} and \figref{parameter_scatter}{}.
    Most of the correlations seen in \figref{parameter_scatter}{} are easy to 
    rationalize. For example, the negative correlation of $E(0)$ and $\al$ 
    tells us that if the number of exposed on day zero is increased, then the 
    probability of testing susceptibles has to decrease to free capacity for 
    the exposed. A similar argument can be made for the negative correlation of 
    $\ka$ and $\al$, which are proportional to the rate of testing exposed 
    and susceptibles, respectively. 
For predicting the future of an
ongoing outbreak, quantification of uncertainties 
due to such parameter correleations is crucial and, thus, should be 
carefully taken into consideration \cite{chowell2017fitting}. 
Because near-real-time
prediction is not our goal here, we focus on the qualitative aspects of our
findings that are robust to parameter uncertainties. 

While \figref{fits}{A} illustrates how an autonomous SEIR model cannot capture
an accelerating epidemic, it also suggest how a nonautonomous SEIR model with a
time-varying transmission parameter $\be (t)$ could have indeed describe the
observed case numbers. One could then have interpreted the accelerating spread
as being due to the seasonal effects such as people spending more time indoors
hence increasing chance of transmission. Another --- somewhat trivial ---
modification to the SEIR model could have been addition of infectious
individuals to the model by hand as a proxy for people bringing the virus from
outside the country through travel as suggested by the Austrian then-Chancellor
Sebastian Kurz who claimed that the sudden increase of the COVID-19 cases during
Fall 2020 was largely due to the Austrians of foreign origin who brought the
virus back from their their countries of origin \cite{thelocal}. It is
conceivable that both of these factors have played some role in the sudden
spread of COVID-19 in Austria in Fall 2020 however they as such do not explain 
the coinciding increase in the test-positive rate. 
As we point out in this study capacity limits in mitigation have played a key role 
in the epidemic acceleration and offer an explanation for the observed events 
including the change in positive rate. 

The methods for estimating the effective reproduction number are usually
believed to be robust against incomplete observations under the assumption that
an approximately same fraction of infections are recorded on consecutive days
\cite{gostic2020practical}. This is also reflected in our initial estimates of
the effective reproduction number, since when the ratio of the number of
infections to that of confirmed cases is constant as in the initial phase of
\figref{fits}{D}, $R_t^{(C)}$ and $R_t^{(I + C)}$ coincide in
\figref{fits}{F}. This,
however, breaks down as soon as the contract tracing limit is reached; after
this point, the effective reproduction number based only on the case data 
systematically underestimates the one that is based on the actual number of
infections. Although the difference between $R_t^{(C)}$ and $R_t^{(I +
C)}$ appear small in \figref{fits}{F}, it should be recalled that this
difference translates into the number of infections exponentially, thus, has a
dramatic real-life consequences such as expected number of hospital admissions. 
Note that the initial overshoot of the effective reproduction number
in \figref{fits}{F} is due to our omission of case data prior 2020-09-01, which
results in an overestimate of the effective reproduction number.

As one should expect, the increase of undetected cases in \figref{fits}{D}
coincide with that of the ratio of positive tests in \figref{fits}{E}, which is
observable during an epidemic. While this information could, and probably
should, be incorporated into statistical methods for estimating $R_t$, we
believe that an increasing positive test ratio is a sufficient reason for a
dramatic intervention such as a lockdown, which is essentially 
inevitable once the contact tracing capacity is exhausted. 

During the second wave of covid-19 in Austria, the policy makers insisted that a
lockdown would be the last option in the country's pandemic response
\cite{pollak2021chronology}. Decreasing lockdown durations for the SEIR model as
shown in \figref{lockdowns}{B} might indeed suggest this as a reasonable
compromise to minimize the number of days during which the economic and social
activities are halted. Note, however, that this behavior changes dramatically in
the SEIRTC model since the uncontrolled spread following the breakdown of
contact tracing makes it progressively harder to reduce the case numbers. This
is why we believe that a steadily increasing ratio of positive tests
necessitates a lockdown. At that point, early action not only saves lives but
also shortens the lockdown duration necessary to regain control.

\bibliography{autoacc}
\section*{Supplementary Material}
\beginsupplement

\subsection*{SEIRTC equations}

The set of ODEs that correspond to the SEIRTC model shown in \figref{models}B read
\eqarr{
    \dot{S} &=& - f + T_{S} \gamma_{T} 
        - \frac{S \beta \left(\rho I_{a} + I_{s}\right)}{N} \label{dotS} \\
    \dot{T_S} &=& f - T_{S} \gamma_{T} \label{dotT_S} \\
    \dot{E} &=& - g E 
            - E \gamma_{E} p (1-g) - E \gamma_{E} \left(1 - p\right) (1-g) 
            + \frac{S \beta \left(\rho I_{a} + I_{s}\right)}{N} \label{dotE} \\
    \dot{T_E} &=& g E - T_{E} \gamma_{T} \label{dotT_E} \\
    \dot{I_a} &=& E \gamma_{E} p (1-g) - I_{a} \gamma_{I} \label{dotI_a} \\
    \dot{I_s} &=& E \gamma_{E} \left(1 - p\right) (1-g) - I_{s} d \gamma_{s} 
        - I_{s} \gamma_{I} \left(1 - d\right) \label{dotI_s} \\
    \dot{T_I} &=& I_{s} d \gamma_{s} - T_{I} \gamma_{T} \label{dotT_I} \\
    \dot{R_u} &=& 
        I_{a} \gamma_{I} + I_{s} \gamma_{I} \left(1 - d\right) \label{dotR_u} \\
    \dot{C} &=& - \gamma_C C + T_{E} \gamma_{T} + T_{I} \gamma_{T} \label{dotC} \\
    \dot{R_k} &=& \gamma_C C    \label{dotR_k}
}
where  $\gamma_C$, $g$, and $f$ are given by \eqref{ga_C}, \eqref{g_ansatz}, 
and \eqref{f_ansatz}, respectively. 

\subsection*{Comparison of fits}

\figref{fitscatter}{} presents a visual comparison of the fits by $SEIR$ and $SEIRTC$ 
models where the recorded cases are plotted against the model predictions. The red 
arrows indicate the maxima of relative prediction errors.

\hfig{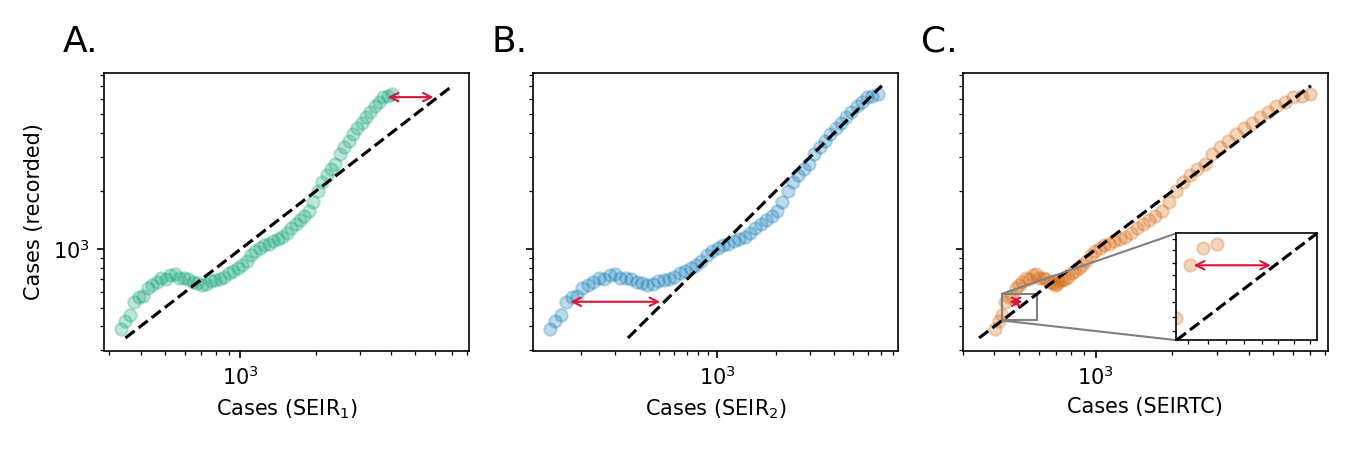}{
    \textbf{Comparison of fits.}
    Scatter plots show the recorded case incidences against the 
    \textbf{A.} SEIR$_1$,  
    \textbf{B.} SEIR$_2$, and 
    \textbf{C.} SEIRTC model predictions.
    Red arrows indicate the locations of largest relative mismatch between 
    the recorded incidence and model predictions which are equal to
    $39.3\%$ (\textbf{A}), $68.8\%$ (\textbf{B}), and $25.0\%$  (\textbf{C}).   
}

As a ``goodness of fit'' test \cite{barlow1993statistics}, we compute the reduced 
chi-squared $\ch_\nu^2 = \ch^2 / \nu$ where $\nu$ is the number of degrees of 
freedom and  
\eq{ch2}{
    \ch^2 = \sum_{i} \frac{\til{X}_i - X_i}{\si_i^2} \,.
}
Here, the sum is over the observations, $\til{X}_i$ is the observed value, such 
as the case numbers on the $i$-th day, $X_i$ is its model prediction and 
$\si_i^2$ is its variance. Since our models only predicts the means and we have 
one observation per day, we do not have a straightforward way of estimating the 
variance of the case and test numbers, which are the observables. We, thus, 
resort to the Poisson assumption, i.e. variance equal to mean, and compute 
\eqref{ch2} assuming $\si_i^2 = X_i$. Finally, we take the number of degrees of 
freedom to be $\nu = n_{obs} - n_{pars}$, where $n_{obs}$ and $n_{pars}$ are 
the number of observations and fit parameters, respectively. Since we fit the 
SEIR models only to the case data, only the case numbers go into the calculation 
of \eqref{ch2}, whereas for SEIRTC we take both case and test data into account. 
Consequently, the number of observations is nearly twice as many in the latter 
case. 

\subsection*{Parameter uncertainties}

Bootstrapping method for quantifying parameter uncertainty consists of generating
synthetic data sets from the best-fit model by adding noise with a presumed
error structure and refitting the model to the newly-generated data
\cite{chowell2017fitting}. As a result, one obtains different sets of parameters
which performs similarly in the training data, which can then be further
analyzed to reveal the parameter correlations. We illustrate this procedure in
\figref{bootstrap}{} where synthetic datasets generated randomly under the
assumption of Poisson-distributed errors are shown along with the predictions
from the best-fit model (\figref{bootstrap}{A,B}). For clarity of illustration
here we only show the first 20 days but the synthetic data were generated for
the entire time interval we consider. After generating a synthetic data, we
fit the model parameters to this newly generated data set, and repeated this
procedure for $100$ realizations. Resulting model predictions differ slightly as
illustrated in \figref{bootstrap}{C,D}. 

\hfig{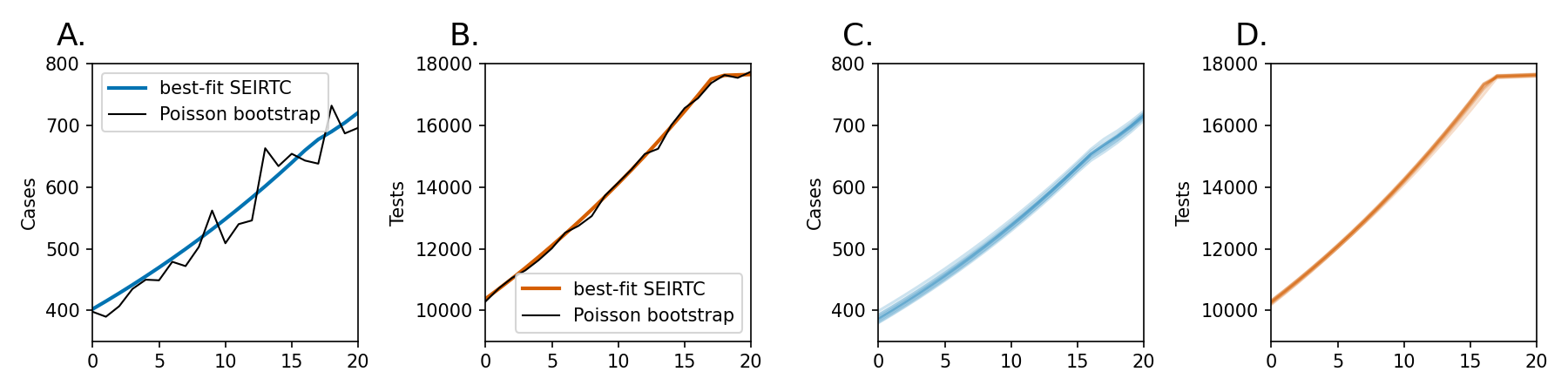}{
    \textbf{Parametric bootstrap method.} 
    Case (\textbf{A}) and tests (\textbf{B}) data (first 20 days) predicted by the 
    best-fit SEIRTC model along with synthetic data generated by assuming 
    Poisson-distributed uncertainties around the means predicted by the SEIRTC 
    model. 
    Case (\textbf{C}) and test (\textbf{D}) data (first 20 days) predicted by the 
    SEIRTC models with 10 different sets of parameters that are obtained by 
    re-fitting the model to synthetic data such as those illustrated in 
    \textbf{A} and \textbf{B}. 
}
 
In order to reveal parameter correlations of the SEIRTC model, we plot the 
best-fit parameters of bootstrap realizations against one another in 
\figref{parameter_scatter}. Here all-but-one of the $100$ bootstrap realizations 
is shown, where we ruled one of the realizations an ``outlier'' for having 
best-fit parameters more than three standard deviations far from the means. 

\hfig{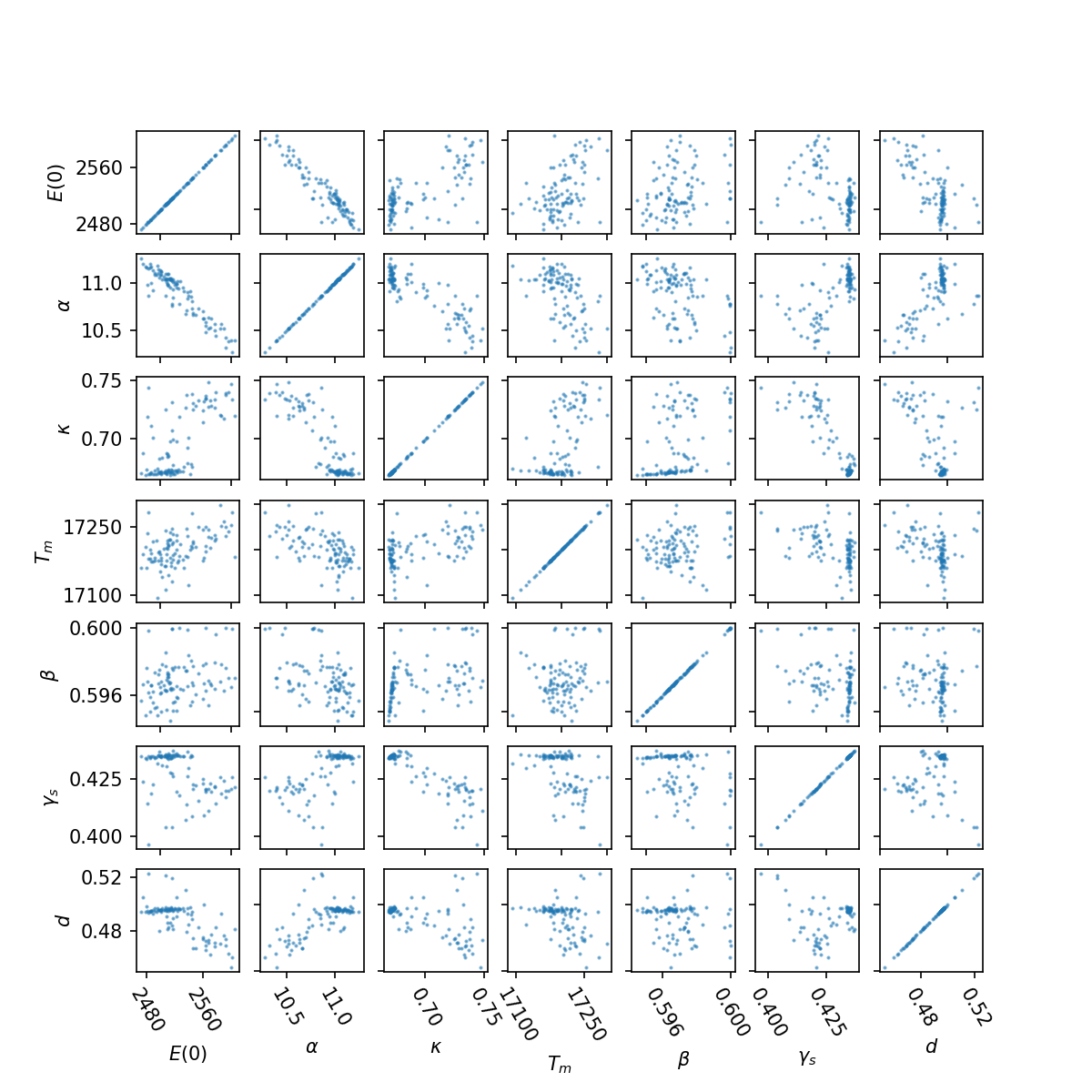}{
    \textbf{Scatters of SEIRTC model parameters.} Each panel shows the scatter of a
    pair of parameters obtained by re-fitting the SEIRTC model to synthetic datasets 
    obtained by the Poisson bootstrap method. 
}

\end{document}